# The effects of Repulsive Biquadratic Interactions in a Blume-Emery-Griffiths Spin-Glass with Competing, Attractive Biquadratic Cross-link Interactions

Daniel P. Snowman

*Department of Physical Sciences, Rhode Island College,*
*Providence, Rhode Island 02908*
(Dated: November 6, 2008)

A BEG hamiltonian is used to model an Ising spin glass with annealed vacancies on a hierarchical lattice. In addition to competing bilinear interactions, repulsive biquadratic interactions on the perimeter of our unit structures compete with attractive cross-link interactions. Ordering and transitions in this system are probed by generating several phase diagrams, using renormalization group methods, for a range of constant $K/J$. A physical interpretation is offered for each sink corresponding to a bulk phase in phase space and critical exponents are calculated for the higher-order transitions.

PACS numbers: 5.70.Fh, 64.60.-i, k75.10.Nr, 5.50.+q

## I. INTRODUCTION

Spin-1 Ising serve as valuable models for materials with fluctuations in magnetization and density. Density in this context refers to the concentration of nonmagnetic impurities, or annealed vacancies in the system. The present investigation using the Blume-Emery-Grittiths (BEG) hamiltonian [1] complete with bilinear ($J_{ij}$), biquadratic ($K_{ij}$) and crystal-field interaction ($\Delta_{ij}$) terms as shown in the Hamiltonian in Eq. (1).

$$-\beta H = \sum_{\langle ij \rangle} J_{ij} s_i s_j + \sum_{\langle ij \rangle} K_{ij} s_i^2 s_j^2 - \sum_{\langle ij \rangle} \Delta_{ij}(s_i^2 + s_j^2)$$
$$\text{with } s_i = 0, \pm 1 \qquad (1)$$

In addition to the normal bilinear interactions, this model includes a crystal-field interaction that directly affects the concentration of nonmagnetic impurities, and a biquadratic coupling that affects the propensity of these impurities to cluster. Each summation in Eq. 1 is over nearest-neighbor $\langle ij \rangle$ pairs of our lattice unit structure including the crystal-field interaction term. The net affect, of this change from sites to bonds for the crystal-field term, being $\Delta$ in eq. (1) is the chemical potential per bond divided by two. The BEG model, originally used to model the superfluid transition in $He^3 - He^4$ mixtures [1], has since been extended to probe the nature of structural glasses [2], microemulsions [3], binary fluids, materials with mobile defects, semiconductor alloys [4], frustrated percolation [5], and a frustrated Ising lattice gas [6, 7].

Phase diagrams and critical phenomenon can be drastically altered due to underlying competing interactions in various Blume-Emery-Griffiths. Many different authors have consider a range of competing interactions using the Blume-Emery-Griffiths model in conjunction with mean-field methods [8?–10] and/or renormalization-group techniques [11–17].

In particular, renormalization-group analysis reveals critical end point and tricritical point topologies linking first and second order phase boundaries for the case with $K > 0$. For repulsive biquadratic coupling ($K < 0$), mean-field calculations revealed two novel phases: one a high-entropy ferrimagnetic phase and the other displaying antiquadrupolar order, see reference [8].

Sellitto et al. [9] focused upon the affects of attractive and repulsive biquadratic interactions upon criticality and resulting phase diagrams using the replica symmetric mean-field approximation with quenched disorder in the bilinear interactions. A spin-glass phase was found with both first and second order transitions from the paramagnetic phase: the crystal-field interaction largely determining the order of the transition. For strong repulsive ($K < 0$) biquadratic interactions, this study also found an antiquadrupolar phase and at lower temperatures, an antiquadrupolar spin-glass phase.

Hierachical lattices have been used with renormalization-group techniques to probe the affects of competing bilinear interactions [10] in a spin-1/2 Ising model, competing bilinear interactions in a BEG system [16], competing biquadratic interactions in a dilute Ising ferromagnet [17], and simultaneous competition between crystal-field and biquadratic interactions in a BEG ferromagnet [15]. Each of these studies employed tuning parameters allowing for the degree of frustration to be decreased, increased or maximized.

The current study complements these earlier works as it employs a BEG hamiltonian to model an Ising spin glass with annealed vacancies on a hierarchical lattice. In addition to competing bilinear interactions, repulsive biquadratic interactions on the perimeter of our unit structures compete with attractive cross-link interactions. Ordering and transitions in this system are probed by generating several phase diagrams, using renormalization group methods, for a range of constant $K/J$. A physical interpretation is offered for each sink corresponding to a bulk phase in phase space and critical exponents are calculated for the higher-order transitions.

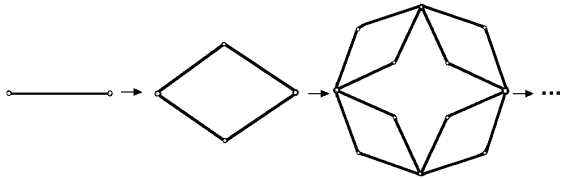

FIG. 1: Repeatedly replacing each nearest-neighbor interaction by the basic generating unit leads to an infinite hierarchical lattice (Berker and Ostlund [25]).

## II. HIERARCHICAL LATTICES AND RENORMALIZATION GROUP THEORY

A basic hierarchical lattice is constructed as shown in Figure 1. The lattice is generated from its basic unit by repeatedly replacing each nearest-neighbor interaction by the basic unit, or generator, itself. The present study has a more complex basic unit, complete with competing cross-link interactions, as shown in Figure 2, and similar to previous studies, see references [18? ]. The renormalization-group solution for a hierarchical model, such as Figures 1 and 2, reverses the construction process. With each renormalization of the system, internal degrees of freedom are removed by summing over all configurations of the innermost sites (represented by solid black dots in Figures 2a and 2b, represented by si, sj in Equation 5).

These lattices are very attractive to use as model systems since the renormalization group recursion relations obtained are exact. As a consequence, these studies can very precisely map phase diagrams and calculate critical exponents. A wide range of complex problems has been studied and better understood using hierarchical lattices. Included amongst these are spin glass [16, 19], frustrated [12, 15–17, 20], random-bond [21], random-field [22], dynamic scaling [23] and directed-path [24] systems. The results of the present study , and those of these previous studies, can be considered approximations into the nature of these systems on more realistic lattices, or, they may be considered exact of these very unique lattices.

Conservation of the systems partition function under renormalization allows for the derivation of the recursion relations relating the interactions at the two length scales. The new effective coupling coefficients $J'$, $K'$, and $\Delta'$ are separated by a distance $l'$ which is $b$ lattice constants in the original system, where b is the length rescaling factor of the renormalization-group transformation.

$$\zeta_{l'}(J', K', \Delta') = \zeta_l(J, K, \Delta) \qquad (2)$$

$$\text{with } l' = bl \qquad (3)$$

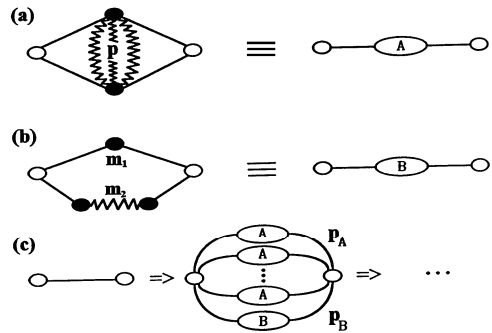

FIG. 2: Construction of the hierarchical lattice. Solid lines represent $(J, K, \Delta)$ nearest-neighbor site interactions, whereas jagged lines represent $(-J, -K, \Delta)$ nearest-neighbor site interactions. An infinite hierarchical lattice is generated from its basic unit by repeatedly replacing each nearest-neighbor interaction by the basic unit itself.

$$\zeta_l = \sum_s \exp[-\beta H] = \sum R_l(s_i, s_j) \qquad (4)$$

$$\text{with } R_l(s_i, s_j) = \sum_{<ij>} \exp[-\beta H] \qquad (5)$$

$$\zeta_{l'} = \sum_{s'} \exp[-\beta H'] = \sum R_l(s_i', s_j') \qquad (6)$$

$$\text{with } R_l(s_i', s_j') = \sum_{<ij>} \exp[J's_i s_j$$
$$+ K's_i^2 s_j^2 - \Delta'(s_i^2 + s_j^2) + \widetilde{G}'] \qquad (7)$$

where $\widetilde{G}'$ is a constant used to calculate the free energy.

Each respective renormalzation-group transformation is calculated by equating individual portions, $R_l(s_i, s_j)$ and $R_l(s_i, s_j)$, of the summation for the partition function at each length scale. These contributions, $R_l(s_i, s_j)$ and $R_l(s_i, s_j)$, correspond to the same fixed configuration of end spins, $s_i, s_j$, at the two different length scales, $l$ and $l$. From these, the relations between interaction strengths at the two length scales, $l$ and $l$, can be derived: $J'(J, K, \Delta), K'(J, K, \Delta)$, and $\Delta'(J, K, \Delta)$. The reader is directed to Section 4 for a derivation of these relations.

Phase diagrams are mapped and transitions characterized using these recursion relations in conjunction with the initial values of $J, K$ and $D$, and the resulting sink(s) of the renormalization-group trajectories.

$$J' = R_J(J, K, \Delta) \qquad (8)$$

$$K' = R_K(J, K, \Delta) \qquad (9)$$

$$\Delta' = R_\Delta(J, K, \Delta) \qquad (10)$$

Each phase has associated with it a corresponding phase sink, see Table I, at which the values of the interactions $(J, K, \Delta)$ have reached a fixed point denoted by $(J*, K*, \Delta*)$. Repeated renormalization will drive the trajectory to one of these sinks. In the vicinity of each fixed point, the system is scale invariant and hence renormalization does not affect the properties of the system as the length scale is increased by a factor of $b$. The fixed points must satisfy the recursion relations such that

$$J^* = R_J(J^*, K^*, \Delta^*) \tag{11}$$

$$K* = R_K(J^*, K^*, \Delta^*) \tag{12}$$

$$\Delta* = R_\Delta(J^*, K^*, \Delta^*) \tag{13}$$

Our calculations have employed a hierarchical lattice generated from a basic unit with two types of components (see Fig. 2a and b), similar to references [12, 15–17, 20]. Each component contains two qualitatively different types of nearest-neighbor sites: those with interactions $(J, K, \Delta)$ and those with interactions $(-J, -K, \Delta)$. The degree of the competition between these two types of interactions, and resulting frustration, is tuned by varying the strength of the cross-link interaction in the unit structure shown Figure 2a. The case of $p = 0$ yields the hierarchical model equivalent [25] to the Migdal-Kadanoff [26, 27] decimation-bond moving scheme in two dimensions. For this study, the crosslink interaction $(p)$, in Fig. 2a, has been chosen (p=4) such that the cross-link interaction leads to maximum frustration in the system.

The end spins are allowed to interact via a second type of component, type B as shown in Figure 2b. This component consists of two different connecting paths: one with $m_1$ pairs of spins all with interactions, $(J, K, \Delta)$, and a second path with $m_2$ pairs of spins, with one nearest-neighbor pair that differs in its nearest neighbor interaction, $(-J, -K, \Delta)$.

Our hierarchical can be viewed as being composed of two different sublattices, distinguished from one another by the type of nearest neighbor interaction, $(J, K, \Delta)$ or $(-J, -K, \Delta)$. Since some sites are connected to neighbors via each type of interaction, it is actually the bonds, rather than the sites, which are associated with each sublattice and must be used to properly interpret ordering and transitions in our system.

The relative number of component structures used in the basic generator for our hierarchical lattice can be varied via two parameters, $p_A$ and $p_B$. This is consistent with previous works that have found a certain level of connectivity required before the full affects of competition are observed; this observation is entirely consistent with other systems, see Kauffman et al. [18], characterized by competing microscopic interactions. The connectivity parameters and competng bilinear and biquadratic nearest neighbor interactions considered in this study parallel those used in previous works [12, 15–17, 20], with $(p, m_1, m_2, p_A, p_B) = (4, 8, 9, 40, 1)$.

## III. CHARACTERIZATION OF PHASE TRANSITIONS

Magnetizations, densities, bilinear and biquadratic nearest neighbor correlations can be calculated by numerically differentiating the free energy with respect to the appropriate variables. The free energy density (dimensionless free energy per bond ), $f$, can be expressed as

$$f = -\frac{\beta F}{N_b} = \sum_{n=1}^{\infty} b^{-nd} G'^n (J^{(n-1)}, K^{(n-1)}, \Delta^{(n-1)}) \tag{14}$$

where $F$ is the Helmhotz free energy and $N_b$ denotes the total number of bonds in the system. The free energy density consists of a sum, over all iterations of the renormalization-group transformation, of the contributions $G'^{(n)}$ to the free energy density due to the degrees of freedom removed during each transformation. Each implementation of the renormalization-group transformation reduces the length scale of the system by a factor of $b$ and the number of spins by a factor of $b^d$.

The free energy density allows us to calculate all thermodynamic quantities. The magnetization, $m \equiv \frac{M}{N_s} = \frac{N_b}{N_s} \frac{\delta f}{\delta H}$, can be calculated by numerically measuring the shift in the free energy density with a small perturbation in the magnetic field, where $N_s$ is the number of sites. Similarly, the density can be calculated by differentiating the free energy density with respect to the crystal field coefficient, $\rho \equiv \frac{N_b}{N_s} \frac{\delta f}{\delta \Delta}$ . Nearest neighbor correlations of the bilinear, $\langle s_i s_j \rangle = \frac{N_b}{N_s} \frac{\delta f}{\delta J}$, and biquadratic, $\langle s_i^2 s_j^2 \rangle = \frac{N_b}{N_s} \frac{\delta f}{\delta K}$ , exchange interactions are also valuable when interpreting the phases and characterizing transitions. Since our unit structure consists of two different types of nearest neighbor interactions, the above thermodynamic quantities have been calculated separately for each type of nearest neighbor pair, $(J, K, \Delta)$ and $(-J, -K, and \Delta)$.

Equipped with the four thermodynamic quantities for each sublattice, discussed above, transitions between the various phases (aka basins of attraction) are characterized. Discontinuities the trademark of first order transitions, whereas second order or continuous transitions exhibit no such discontinuities. Exact knowledge of the renormalized coupling coefficients allow us to precisely calculate critical scaling exponents as higher-order transitions are encountered in our parameter space. The next section explores this in greater detail.

## IV.   RECURSION RELATIONS

By equating the contributions to the partition function from the two length scales, for each fixed end spin configuration, we can write the following equalities for the type A structure shown in Figure 2a.

$$R_l[1,1] = \exp[-4\Delta] + 2\exp[-2J + 2K - \Delta(6+p)]$$
$$+ 2\exp[2J + 2K - \Delta(6+p)] + \exp[J(-4-p) + K(4-p) - \Delta(8+2p)]$$
$$+ \exp[J(4-p) + K(4-p) - \Delta(8+2p)] + 2\exp[K(4-p) + Jp - \Delta(8+2p)]$$
$$= \exp[J' + K' - 2\Delta' + \tilde{G}] = R_{l'}[1,1]$$

$$(15)$$

$$R_l[1,0] = \exp[-2\Delta] + 2\exp[-J + K - \Delta(4+p)] + 2\exp[J + K - \Delta(4+p)]$$
$$+ \exp[J(-2-p) + K(2-p) - \Delta(6+2p)] + \exp[J(2-p) + K(2-p) - \Delta(6+2p)]$$
$$+ 2\exp[K(2-p) + Jp - \Delta(6+2p)]$$
$$= \exp[-\Delta' + \tilde{G}] = R_{l'}[1,0]$$

$$(16)$$

$$R_l[1,-1] = \exp[-4\Delta] + 4\exp[2K - \Delta(6+p)]$$
$$+ 2\exp[K(4-p) - Jp - \Delta(8+2p)] + 2\exp[K(4-p) + Jp - \Delta(8+2p)]$$
$$= \exp[-J' + K' - 2\Delta' + \tilde{G}] = R_{l'}[1,-1]$$

$$(17)$$

$$R_l[0,0] = \exp[0] + 4\exp[-\Delta(2+p)] + 2\exp[-Jp - Kp - \Delta(4+2p)]$$
$$+ 2\exp[Jp - Kp - \Delta(4+2p)] = \exp[\tilde{G}] = R_{l'}[0,0]$$

$$(18)$$

Using the relationships above (Eqs. 15-18) we can derive expressions relating the coupling coefficients between the two length scales for the type A unit structure.

$$J'_A = \frac{1}{2}\log\frac{R_{l'}(1,1)}{R_{l'}(1,-1)} \tag{19}$$

$$K'_A = \frac{1}{2}\log\frac{R_{l'}(1,1)R_{l'}(1,-1)R_{l'}^2(0,0)}{R_{l'}^4(1,0)} \tag{20}$$

$$\Delta'_A = \log\frac{R_{l'}(0,0)}{R_{l'}(1,0)} \tag{21}$$

$$\widetilde{G}'_A = \log R_{l'}(0,0) \tag{22}$$

The recursion relations for simpler (type B) unit structures have the same form as in Eqs. 19-22, but the expressions (Eqs.15-18) for the corresponding Rl(si,sj) differ. Combining the recursion relationships for both types of structures (type A and type B as shown in Fig. 2), the renormalization relationships become

$$J' = p_A J'_A + p_B J'_B \tag{23}$$

$$K' = p_A K'_A + p_B K'_B \tag{24}$$

$$\Delta' = p_A \Delta'_A + p_B \Delta'_B \tag{25}$$

The exact recursion relations above can be used to calculate critical exponents by linearizing the recursion relations in the vicinity of the second-order transition under investigation. That is,

$$J' - J^* = T_{JJ}(J - J^*) + T_{JK}(K - K^*)$$
$$+ T_{J\Delta}(\Delta - \Delta^*), \tag{26}$$

| Phase | Sink | Characteristics |
|---|---|---|
| Paramagnetic I | $J \to 0$ | Dense Sublattice I |
| | $K \to 0$ | Dilute Sublattice II |
| | $\Delta \to -\infty$ | |
| Spin Glass I | $J \to chaotic$ | Dense Sublattice I |
| | $K \to chaotic$ | Dilute Sublattice II |
| | $\Delta \to -\infty$ | |
| Paramagnetic II | $J \to 0$ | Dense Sublattice II |
| | $K \to 0$ | Dilute Sublattice I |
| | $\Delta \to +\infty$ | |

TABLE I: Phases and Corresponding Sinks

$$K^{'} - K^* = T_{KJ}(J - J^*) + T_{KK}(K - K^*) \\ + T_{K\Delta}(\Delta - \Delta^*), \quad (27)$$

$$\Delta^{'} - \Delta^* = T_{\Delta J}(J - J^*) + T_{\Delta K}(K - K^*) \\ + T_{\Delta\Delta}(\Delta - \Delta^*), \quad (28)$$

,

where $T_{JJ} = \frac{\delta J^{'}}{\delta J}$, $T_{KJ} = \frac{\delta K^{'}}{\delta J}$, etc. and are evaluated at the fixed point in question. The critical relations above can be represented as a recursion matrix, with elements $T_{XY}$ and eigenvalues of the form

$$\Lambda_l = b^{y_l} \quad (29)$$

where $b$ is the length rescaling factor (in our case $b = 2$) and $y_l$ represents the corresponding critical exponent for the $l_{th}$ eigenvalue.

## V. RESULTS

The results below detail our investigation into the affects of varying temperature( $1/J$) and vacancy of nonmagnetic impurities ( $\Delta/J$) in a system characterized by competing biquadratic and bilinear interactions. A series of planes of constant biquadratic interactions were considered, with exhaustive analysis of the resulting renormalization-group trajectories in parameter space. Three unique basins of attraction were found, each sharing a renormalization-group trajectory that flows to a common sink (see Table I), and each corresponding to a unique phase in parameter space.

Our exploration has yielded two unique paramagnetic phases, Paramagnetic I and Paramagnetic II, in addition to a Spin Glass I phase. It is helpful if we split our lattice into two sublattices: type I and type II. The type I sublattice consists of those sites interacting via coupling coefficients $(J, K, \Delta)$ and, our type II sublattice consists of those sites interacting via $(-J, -K, \Delta)$ interactions. The two types of phases, type I and type II, fundamentally differ in the distribution of magnetic species. The renormalization-group flow for the crystal-field interaction determines the density on each sublattice. Type I phases correspond to a crystal-field interaction term flowing to -, whereas a type II phase correspond to a flow to +. A type I phase has a large density of magnetic species on sublattice I and a dilute population on sublattice II, and vice versa. The bilinear and biquadratic interactions flow to zero in both paramagnetic phases. The spin glass phase, Spin Glass I, is has a distribution of magnetic species similar to that of Paramagnetic I but it also has nonzero flows for the bilinear and biquadratic interaction indicating the presence of magnetic and spatial ordering.

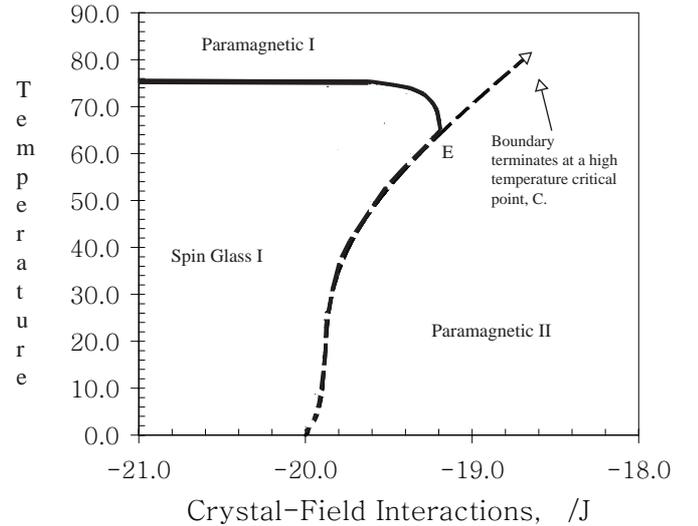

FIG. 3: Parameter space, with K/J = -20 and $(p, m_1, m_2, p_A, p_B) = (4, 8, 9, 40, 1)$, depicting different basins of attraction and associated phases with critical endpoint (E) and critical point (C). Solid lines represent second-order transitions, whereas dashed lines represent first-order transitions.

We focus our attention first on the plane of constant biquadratic coupling with K/J = -20. In this plane we find a Paramagnetic I, Paramagnetic II and Spin Glass I phase. The type I phases found at the more negative values of the crystal-field interaction. This is consistent with the expectation, since negative $\Delta/J$ corresponds to a larger concentration of occupied sites on sublattice I and a correspondingly dilute sublattice II.

The two paramagnetic phases are separated by a first order phase boundary at high temperatures. This phase boundary can be traversed via a changing temperature and/or crystal field interaction ($\Delta/J$). This phase boundary terminates at a high temperature critical point C. At temperatures above this critical point it is possible for the system to pass smoothly between the two phases as in the standard liquid-gas phase diagram.

From the Paramagnetic I phase, a decrease in the temperature forces the system through an ordering transition to the Spin Glass I phase. This transition is second-order

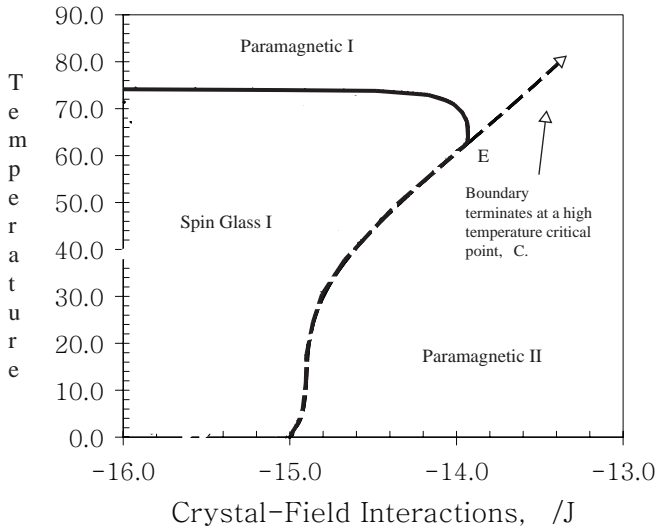

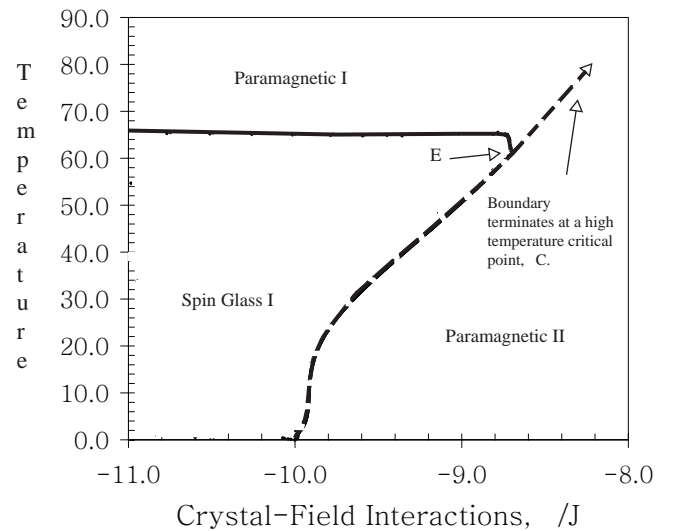

FIG. 4: Parameter space, with K/J = -15 and $(p, m_1, m_2, p_A, p_B) = (4, 8, 9, 40, 1)$, depicting different basins of attraction and associated phases with critical endpoint (E) and critical point (C). Solid lines represent second-order transitions, whereas dashed lines represent first-order transitions.

FIG. 5: Parameter space, with K/J = -10 and $(p, m_1, m_2, p_A, p_B) = (4, 8, 9, 40, 1)$, depicting different basins of attraction and associated phases with critical endpoint (E). Solid lines represent second-order transitions, whereas dashed lines represent first-order transitions.

in nature and occurs over a range of crystal-field interactions, or concentrations, at approximately $1/J \approx 75$. This critical line terminates at critical endpoint E upon its intersection with the line of first order transitions similar to the topology observed by Hoston and Berker [7] for the case of uniform J, K and $\Delta$ with K/J = 5, using mean-field theory

From within the Spin Glass I phase, an increase in the concentration of nonmagnetic impurities on sublattice I forces the system to disorder. This increase in the crystal-field interaction drives the system to the Paramagnetic II state via a first order transition. Note, the intersection of the line of first order transition at zero temperature occurs at $\Delta/J = K/J$.

An increase in the clustering bias, or biquadratic coupling, to K/J = -15 results in a shift of the first order phase boundary to larger $\Delta/J$. The topology of the critical endpoint (E), critical point (C) phase diagrams remains intact. The zero temperature intersection of the first-order boundary has shifted to $\Delta/J = -15$.

An increase to K/J = -10, results in the second-order Paramagnetic I/Spin Glass I phase boundary curving less, whereas the first order phase boundary remains curved across the entire span of temperatures depicted. The location of the critical endpoint (E) remains at approximately the same temperature, $1/J \approx 60$.

In the K/J = -5 plane we see evidence, as in the previous planes considered, of a Spin Glass I phase that can disorder to paramagnetic states three different ways. An increase in temperature disorders the spin glass phase to the Paramagnetic I phase. A decrease in temperature drives the system to disorder to the Paramagnetic

II phase via a first order transition. And, finally, an increase in the crystal-field interaction drives the system to the Paramagnetic II phase as the concentration of nonmagnetic impurities is increased on sublattice I.

In the last plane of constant biquadratic coupling considered, K/J = -1, we find a phase diagram with a line of first order transitions that is nearly vertical. Thus, the Spin Glass I phase can no longer disorder as a result of a decreasing temperature in the system. However, the critical endpoint (E)/critical point (C) topology remains qualitatively unchanged.

In the current investigation, the nature of critical scaling on the secon-order phase boundary separating the Paramagnetic I phase from the Spin Glass I phase at high temperatures has been probed. Linearizing the recursion relations while maintaining four scaling fields, associated with J, K, $\Delta$ and H (as discussed in Section 4) yields a recursion matrix. Calculation of the eigenvalues allow us to determine critical scaling exponents. Conducting this analysis for the critical line separating the Paramagnetic I and Spin Glass I phases yields: two relevant eigenvalues, $\Lambda_1 = 2.11$ and $\Lambda_2 = 1.29$, corresponding to critical scaling exponents of $y_1 = 1.08$ and $y_2 = 0.367$, respectively; and, two irrelevant eigenvalues with $\Lambda_3 = 0.037$ and $\Lambda_4 = -0.022$.

## VI. SUMMARY

In summary, a BEG hamiltonian is used to model an Ising spin glass with annealed vacancies on a hierarchi-cal lattice. In addition to competing bilinear interactions, repulsive biquadratic interactions on the perimeter of our unit structures compete with attractive cross-link interactions. Ordering and transitions in this system are probed by generating several phase diagrams, using renormalization group methods, for a range of constant $K/J$. A physical interpretation is offered for each sink corresponding to a bulk phase in phase space and critical exponents are calculated for the higher-order transitions.

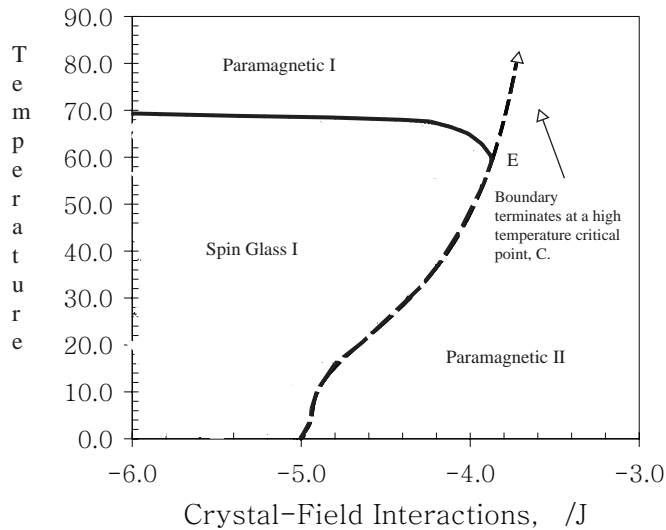

FIG. 6: Parameter space, with K/J = -5 and $(p, m_1, m_2, p_A, p_B) = (4, 8, 9, 40, 1)$, depicting different basins of attraction and associated phases with critical endpoint (E) and critical point (C). Solid lines represent second-order transitions, whereas dashed lines represent first-order transitions.

In our investigations of several planes of constant, repulsive biquadratic interactions three phases were identified: Paramagnetic I, Paramagnetic II and Spin Glass I. Each phase distinguished by a unique sink for its renormalization group trajectory. The Spin Glass I phase disordering to its Paramagnetic I counterpart, at high temperatures, via a second order transiton. This line of criticality terminates at a critical endpoint E shared with the line of first order transitions. Spin glass systems and the affects upon ordering of nonmagnetic impurities may potentially be better understood as a result of these investigations.

[1] M. Blume, V. Emery, and R. Griffiths, Phys. Rev. A **4**, 1071 (1971).

[2] K. T.R. and T. D., Phys. Rev. B **36**, 5388 (1987).

[3] M. Schick and W.-H. Shih, Phys. Rev. B **34**, 1797 (1986).

[4] K. Newman and J. D. Dow, Phys. Rev. B **27**, 7495 (1983).

[5] A. Coniglio, J. Phys. IV France **3**, C1 (1993).

[6] J. Arenzon, M. Nicodemi, and M. Sellitto, J. Phys. I France **6**, 1143 (1996).

[7] M. Nicodemi and A. Coniglio, J. Phys. A **30**, L187 (1997).

[8] W. Hoston and A. Berker, Phys. Rev. Lett **67**, 1027 (1991).

[9] M. Sellitto, M. Nicodemi, and J. Arenzon, J. Phys. I France **7**, 945 (1997).

[10] D. P. Snowman, PhD dissertation, University of Maine, Orono, ME (1995).

[11] A. Berker and M. Wortis, Phys. Rev. B **14**, 4946 (1976).

[12] A. B. S.R. McKay, Phys. Rev. Lett. **48**, 767 (1982).

[13] N. S. Branco, arXiv:cond-mat/9904082v1.

[14] N. S. Branco and B. M. Boechat, arXiv:cond-mat/9708043v3.

[15] D. Snowman, J. Magn. Magn. Mater. **314**, 69 (2007).

[16] D. Snowman, J. Magn. Magn. Mater. **320**, 1622 (2008).

[17] D. Snowman, Phys. Rev. E **77**, 041112 (2008).

[18] M. Kaufman and R. Griffiths, Phys. Rev. B **24**, 496 (1981).

[19] G. Migliorini and A. Berker, Phys. Rev. B **57**, 426 (1998).

[20] A. B. S.R. McKay and S. Kirkpatrick, J. Appl. Phys. **53**, 7974 (1982).

[21] D. Andelman and A. Berker, Phys. Rev. B **29**, 2630 (1984).

[22] A. B. A. Falicov and S. McKay, Phys. Rev. B **51**, 8266 (1995).

[23] R. Stinchcombe and A. Maggs, J. Phys. A **19**, 1949 (1986).

[24] R. da Silveira and J. Bouchaud, Phys. Rev. Lett. **93**, 015901 (2004).

[25] A. Berker and S. Ostlund, J. Phys. C **12**, 4961 (1979).

[26] A. Migdal, Zh. Eksp. Teor. Fiz **69**, 1457 (1975).

[27] L. Kadanoff, Ann Phys. New York **100**, 359 (1976).

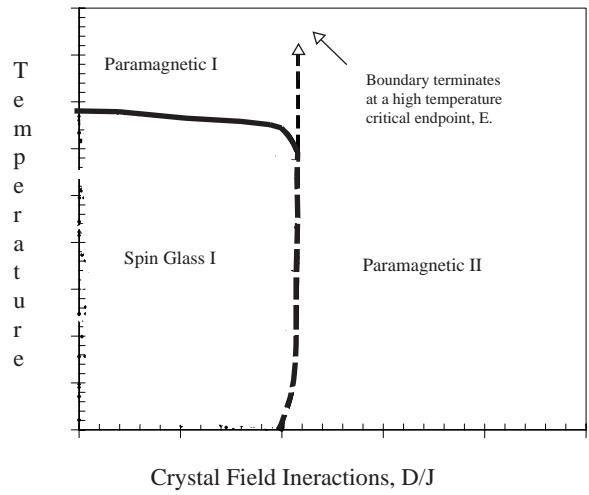

FIG. 7: Parameter space, with K/J = -1 and $(p, m_1, m_2, p_A, p_B) = (4, 8, 9, 40, 1)$, depicting different basins of attraction and associated phases with critical endpoint (E) and critical point (C). Solid lines represent second-order transitions, whereas dashed lines represent first-order transitions.

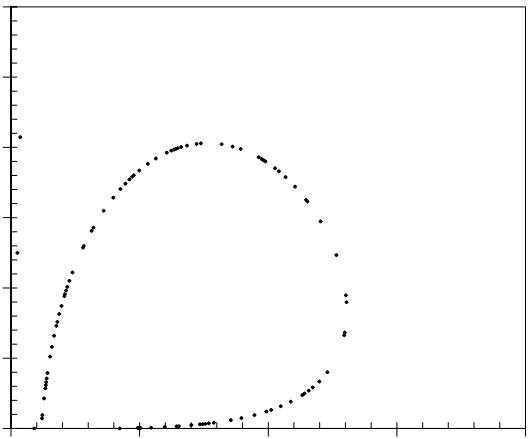

FIG. 8: Recursion relations in bulk of glassy phase for parameters....